\documentclass[twocolumn,prl,superscriptaddress]{revtex4}
\usepackage[colorlinks,linkcolor=blue,anchorcolor=blue,citecolor=blue,filecolor=blue,menucolor=blue,runcolor=blue,urlcolor=blue,frenchlinks=blue]{hyperref}
\usepackage{amsmath}
\usepackage{graphicx}
\usepackage{amssymb}
\usepackage{dcolumn}
\usepackage{mathrsfs}
\usepackage{bm}
\usepackage{color}
\begin{document}

\title{Fuzzy-processing quantum computation}

\author{Yan-Xiong Du}
\email{yanxiongdu@m.scnu.edu.cn}
\affiliation {Key Laboratory of Atomic and Subatomic Structure and Quantum Control (Ministry of Education), Guangdong Basic Research Center of Excellence for Structure and Fundamental Interactions of Matter, School of Physics, South China Normal University, Guangzhou 510006, China}

\affiliation {Guangdong Provincial Key Laboratory of Quantum Engineering and Quantum Materials, Guangdong-Hong Kong Joint Laboratory of Quantum Matter, Frontier Research Institute for Physics, South China Normal University, Guangzhou 510006, China}


\begin{abstract}
Quantum computation has attracted numerous attentions and develops rapidly in the recent decades. To against the decoherence and the control errors upon the qubits, quantum error corrections are adopted. Such approaches require lots of redundant qubits, accurate measurement and timely feedback. Here we investigate a new framework of quantum computation that is associated with fuzzy processing. It will benefit significantly from three aspects: the fuzzy recognition of qubit states reduce the required gate fidelity; the fuzzy encoding encodes the information of the qubits into a distribution of probability, suppressing the fluctuations in the output of long quantum circuits; the fuzzy feedback offers a more efficient way to control the qubits when precision information of quantum states are absent. Furthermore, the fuzzy processing can be integrated into quantum error correction, eliminating the need for immediate correction operations. The proposed scheme will be fairly suitable for the solution of decision problems, which has significant applications in the optimization problems and control problems.
\end{abstract}

\maketitle

\textit{Introduction.}--- Quantum computer use quantum states superposition and entanglement to accelerate computation. It has shown great potentials in dealing with optimization problem, prime factorization, machine learning and so on. Quantum error correction (QEC) is an essential procedure to realize large-scale quantum computation by eliminating errors and resisting decoherence \cite{Shor1995,Steane1996,Dennis2002,Terhal2015}. Proof-of-principle progresses of QEC have been achieved in neutral-atom \cite{Bluvstein2023,Bluvstein2025,Atom2026}, superconductor \cite{Google2023,Sivak2023,Google2025,He2025,Sun2025}, trapped ions \cite{Hong2024,Pogorelov2025,Paetznick2026,Tham2026}, quantum dots \cite{Takeda2022,Riggelen2022,Zhang2026} and many other systems \cite{Selim2024,Larsen2025,Guo2025}. To further improve the performance of QEC, several technological problems should be solved \cite{Terhal2015,Fowler2012,Roffe2019,Campbell2024}, for instance, the errors come from crosstalk, states leakage and measurement; feedback control of error corrections; decoherence of qubits, and so on. More efficient QEC schemes are also proposed. As an example, the family of quantum low-density parity-check codes will use much less physical qubits to encode logical qubits \cite{Breuckmann2021,Bravyi2024,Poole2025,Pecorari2025}. Such schemes will be implemented at the prize of much more non-local manipulations. At present, quantum computations are still undergoing a sharp iterative upgrade.

In fact, the quantum computers are complex physical systems that contain nonlinear interactions, imperfect controls and detections. As the scale of the systems getting larger and larger, one may seek some new forms of manipulations. Such situations are quite similar to fuzzy control which deals with analog signals. There the complex configuration of the control systems are hard to be modeled, furthermore, it becomes uncertain due to the coupling to the environment. Such control process would be stable for long time once we fuzzify the inputs and outputs properly. Over the past decades, the fuzzy control has been applied to intelligent control (e. g., energy harvesting, ambient conditioning systems, robotics and
autonomous systems), pattern recognition and categorization, and so on \cite{Lam2018,Tang2024,Singh2025}, outperforming the traditional linear controller. Prior to this, some attempts have been made to combine fuzzy logic and quantum computation, for example, performing Kaufmann's addition of fuzzy numbers in quantum circuits \cite{Hooghe2004}; designing fuzzy estimators for quantum control with uncertainties \cite{Chen2012}. Nevertheless, the aforementioned works have not solved the major problems in quantum computations.

In this letter, we propose to realize quantum computation with fuzzy processing. Here we only focus on the decision problems, that is, quantum circuit whose output is either yes or no (0\textbackslash 1). It is known that the optimization problems or search problems, which are Non-deterministic Polynomial problems, can decomposed into a group of decision problems (wether the problem has a solution within a certain range of parameters). The fuzzy logic is suggested to apply to the entire process of decision problems in quantum algorithms. Through adopting fuzzy recognition to identify the distribution of the output states but not a specific value, the required fidelity of quantum gate is reduced. The fuzzy encoding of the input states will help to narrow the distribution of the outputs, in a sufficiently large circuit under the random noises. The fuzzy logic allows one to realize effective control over quantum states even in the absence of precise information about the system, permitting quantum circuit with $10^5$-level amount of quantum gates under an average fidelity 0.999. Finally, the combination of QEC and fuzzy processing is investigated. Within the framework of decision problems and fuzzy encoding, the timely error correction at each computational layer is abandoned. This advantage can significantly reduce the difficulty of realizing quantum computing.

\textit{Brief review of fuzzy mathematics.}--- Here we review the basic concept of fuzzy mathematics briefly. In the seminal paper of L. A. Zadeh \cite{Zadeh1965}, fuzzy set theory is introduced and extends classical set theory by permitting partial membership. Unlike crisp sets where elements either belong or do not, a fuzzy set allows degrees of membership ranging from 0 to 1. It effectively models the uncertainty and vagueness inherent in human reasoning and complex systems \cite{Chen2000,Tanaka1992}.

Let $\chi$ be the universal set, a fuzzy set $\mathcal{A}$ in $\chi$ is a set of ordered pairs $\mathcal{A}:=\{(x,\mu_\mathcal{A}(x)), x\in\chi\}$. $\mu_\mathcal{A}:\chi\rightarrow[0, 1]$ is called the membership function which characterize the grade of membership of $x$ in $\mathcal{A}$. When $\mu_\mathcal{A}:\chi\rightarrow\{0, 1\}$, $\mathcal{A}$ returns to a crisp set. Function $\mu_\mathcal{A}$ can be chosen to be triangular, trapezoidal, Gaussian, or others depended on the systems \cite{Chen2000}. A vector $\mathcal{V}$ is called a fuzzy vector if all the elements belong to the interval $[0, 1]$, i. e., $\mathcal{V}(x)=(\mathcal{V}_i)=(\mu_i(x))$ represents the membership functions of $x$ belong to fuzzy set $\mathcal{A}_i$. The fuzzy matrix can also be introduced to transform the fuzzy vector to another one, where the elements of fuzzy matrix are all belong to the interval $[0, 1]$. Given a fuzzy matrix $\mathcal{M}=(\mathcal{M}_{ij})_{m\times n}$ and a fuzzy vector $\mathcal{V}=(\mathcal{V}_j)$, the product of $\mathcal{M}$ and $\mathcal{V}$ is given by $\mathcal{V}'=\mathcal{M}\circ \mathcal{V}$ with
\begin{equation}
\mathcal{V}'_k=\vee_{j=1}^{m}(\mathcal{M}_{kj}\wedge\mathcal{V}_j).
\end{equation}
Here $\wedge$ represents taking the minimum element from the group and $\vee$ represents taking the maximum one. Through fuzzifying the input of the system with a fuzzy vector and retrieve the output with multiplication (1), the fuzzy control can be realized. It has been demonstrated that the fuzzy control can handle ill-defined and complex nonlinear system. The explanation of such stabilities is describe by the dynamics of nonlinear system as an average weighted sum of some local linear subsystems where the weights characterized by membership functions measure the contribution made by each \cite{Chen1995,Tanaka2001}.

\textit{Fuzzy recognition of quantum states.}--- Applying fuzzy logic to quantum state discrimination is a natural choice in the framework of decision problems. Here we may consider a circuit which is constructed by a unitary operation $U(w,\lambda)$ (that may contain a group operations) and single qubit $|\psi\rangle$ with computational basis $|0\rangle=(1,0)^{\mathbf{T}}, |1\rangle=(0,1)^{\mathbf{T}}$, ${\mathbf{T}}$ denotes the transpose operation. $w$ is the control parameter and $\lambda$ characterizes the intensity of control noise. We set  $U(0,0)=I$ and $U(1,0)=\sigma_x$ through adjusting the coupling strength or the interaction time of the system, $I$ is the identity matrix and $\sigma_x$ (also $\sigma_{y,z}$) is one of the Pauli Matrix. Given that the initial state is $|\psi\rangle_\mathrm{i}=|\alpha\rangle$, we will obtain the final state with
\begin{equation}
|\psi\rangle_\mathrm{f}=U(w,\lambda)|\psi\rangle_\mathrm{i}.
\end{equation}
The measurement results can be characterized by quantity $K(\lambda)=|\langle\alpha|\psi\rangle_\mathrm{f}|^2$. In the ideal case with $\lambda=0$, $K=1$ due to $U=I$ and $K=0$ as $U=\sigma_x$. One will concern the noise intensity beyond which the output $K$ is no longer distinguishable, in spite of the measurement precision. Actually, in the framework of fuzzy logic, it can be obtained that
\begin{equation}
|\psi\rangle_\mathrm{f}=
\begin{cases}
|\alpha\rangle & \text{if } K(\lambda)>0.5+\epsilon \\
|\beta\rangle & \text{if } K(\lambda)>0.5-\epsilon
\end{cases},
\end{equation}
where $\langle\alpha|\beta\rangle=0$, $\epsilon$ is the fluctuation of the output that due to random noises. Therefore, the quantum gates of realizing $U$ need not to be perfect.

\begin{figure}[ptb]
\begin{center}
\includegraphics[width=8.7cm]{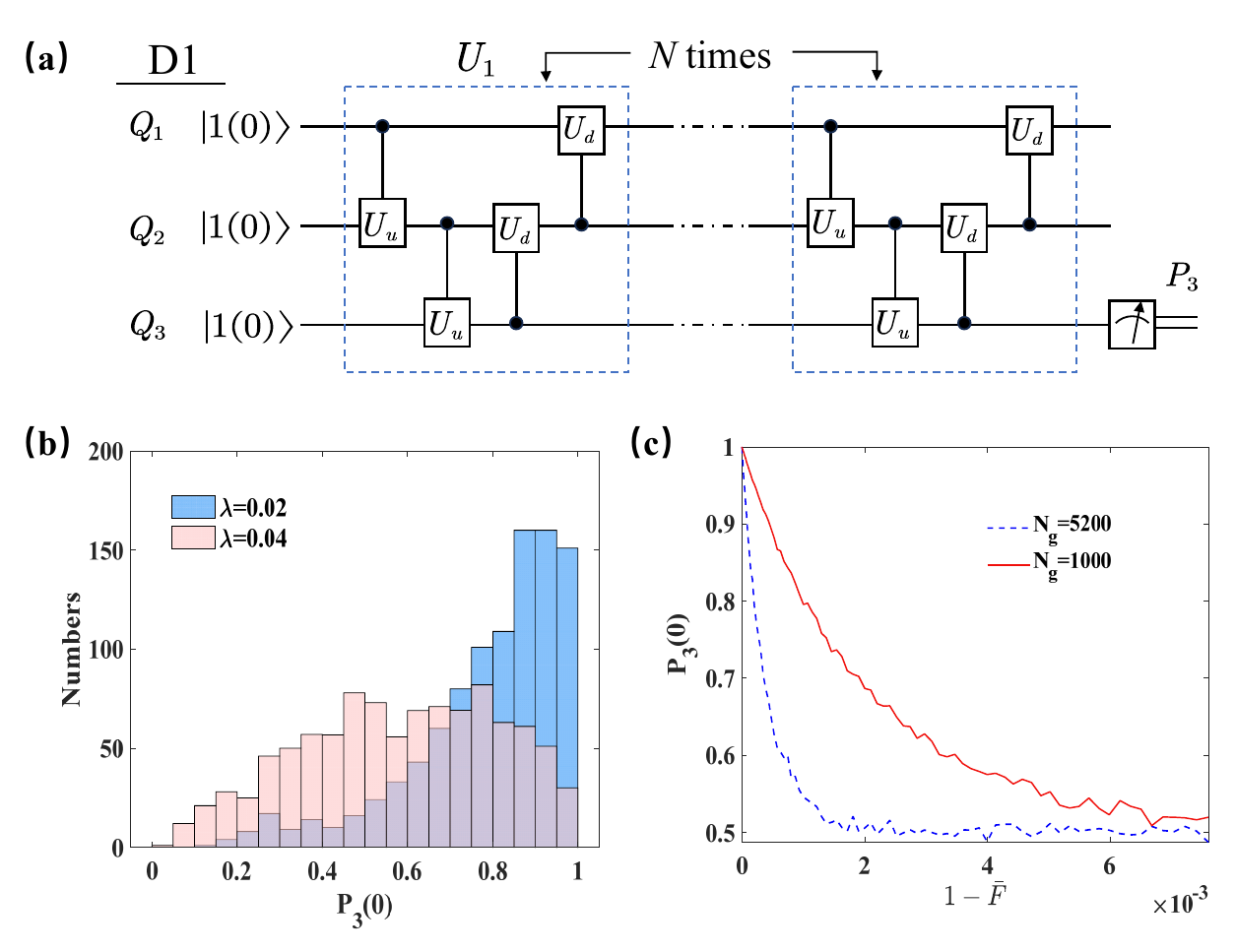}
\caption{
Fuzzy recognition in quantum circuit. (a) The tested quantum circuit D1. (b) The distribution of $P_3(0)$ with different intensity of noise $\lambda=0.02$ and $\lambda=0.04$, respectively. (c) The output $P_3(0)$ versus infidelity $1-\bar{F}$ with different amount of quantum gates $N_g$.
}
\end{center}
\end{figure}

As shown in Fig. 1(a), we investigate the improvement of using fuzzy recognition \cite{Note1}. A quantum circuit D1 of three qubits $Q_{1,2,3}$ is introduced, where $Q_{1,2}$ are the control qubits and $Q_3$ is the target qubit. Operation $U_1$ is repeated by $N$ times with $U_1=U_d^{(2,1)}U_d^{(3,2)}U_u^{(2,3)}U_u^{(1,2)}$,
\begin{equation}
\begin{aligned}
U_u=|0\rangle\langle0|\otimes I+|1\rangle\langle1|\otimes U_{\theta},\\
U_d=I\otimes|0\rangle\langle0|+U_{\theta}\otimes|1\rangle\langle1|.\\
\end{aligned}
\end{equation}
$I$ is the identity matrix and $U_\theta=e^{i\theta\sigma_x}$. $U_u^{(1,2)}=U_u\otimes I$, $U_u^{(2,3)}=I\otimes U_u$, $U_d^{(3,2)}=I\otimes U_d$ and $U_d^{(2,1)}=U_d\otimes I$. It can be checked that through setting $N=20N_k$ ($N_k=1, 2, 3, \cdots$) and $\theta=\theta_0=\pi/2$, $Q_3$ will be flipped when $Q_1Q_2=|11\rangle$ and $Q_3$ will remain unchanged when $Q_1Q_2=|01\rangle$. In Fig. 1(b), the output of D1 against random noise with initial state $|111\rangle$ is discussed, $N=1300$. The random noise is introduced by $\theta=\theta_0(1+\lambda rand)$, $rand$ is an array of random numbers in $[-1, 1]$ with a mean of zero. The output of $Q_3$ is characterized by $P_3(0)=\langle I\otimes I\otimes (1+\sigma_z)\rangle/2$. In the ideal case, $P_3(0)=1$. Due to the random noise upon the quantum gates, the output $P_3(0)$ will spread. We exhibit two distribution of $P_3(0)$ with $\lambda=0.02$ and $\lambda=0.04$, respectively. The bar chart results are derived from 1,000 simulations. Obviously, the increase of the intensity of the random noise will exacerbate the spread of the distribution. In Fig. 1(c), the relationship between $P_3(0)$ and infidelity $1-\bar{F}$ is discussed. The average fidelity of quantum gates over $N$ rounds is defined by $\bar{F}=\overline{\mathbf{Min}\{|\mathbf{Tr}({U'}^{(k)}_{\mathrm{id}}U^{(k)}_\mathrm{g})|^2/4\}}$, $U^{(k)}_\mathrm{g}$ is the actual evolution operator  and $U^{(k)}_{\mathrm{id}}$ is the ideal one of the $k$-th gate, $\mathbf{Tr}$ denotes the trace operation. $\mathbf{Min}$ function returns the minimal value at each round. Data in Fig. 1(c) has been averaged for 1000 times \cite{limit}. When the total amount of quantum gates is $N_g=4N=1000$, $P_3(0)$ can be distinguished with $\bar{F}>0.995$; Meanwhile, when $N_g=5200$, $P_3(0)$ is distinguishable with $\bar{F}>0.999$.

\begin{figure}[ptb]
\begin{center}
\includegraphics[width=8.7cm]{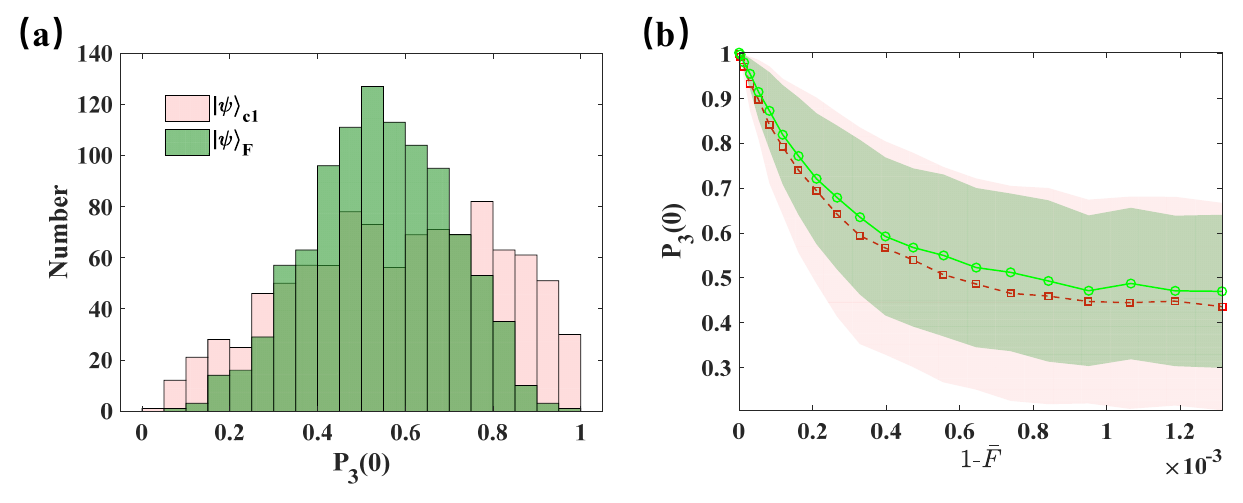}
\caption{
Fuzzy encoding in circuit D1. (a) The distribution of $P_3(0)$ with different input states $|\psi\rangle_{\mathrm{c1}}$ and $|\psi\rangle_{\mathrm{F}}$. (b) The output $P_3(0)$ versus infidelity $1-\bar{F}$ with different input states. Green-solid line with circles: $|\psi\rangle_{\mathrm{F}}$; red-dashed line with squares: $|\psi\rangle_{\mathrm{c1}}$.
}
\end{center}
\end{figure}

\textit{Fuzzy encoding of qubits.}--- We introduce fuzzy encoding of qubits to further improve the performance of fuzzy processing. Different from the ideal input with single computational basis $|\varphi\rangle_\mathrm{c}$ exactly, the fuzzy state input will be characterized by a distribution as
\begin{equation}
|\psi\rangle_\mathrm{F}=\sum_\varphi f(\varphi)|\varphi\rangle_\mathrm{c}.
\end{equation}
In Fig. 2(a), the performance of single state (fuzzy state) input in circuit D1 with $N=1300$ is discussed. The intensity of the noise is set to be $\lambda=0.04$ and the data comes from 1000 times statistics. The single state input is $|\varphi\rangle_{\mathrm{c1}}=|111\rangle$ while the fuzzy state input $|\psi\rangle_\mathrm{F}=|\widetilde{111}\rangle=\sqrt{0.6}|111\rangle+\sqrt{0.2}|110\rangle+\sqrt{0.2}|101\rangle$. The fuzzy state can be prepared by $|\psi\rangle_\mathrm{F}=U_{\mathrm{F}}|\varphi\rangle_{\mathrm{c1}}, U_{\mathrm{F}}=U_\mathrm{s}\mathrm{CU}_{\pi/4}^{(2,3)}\mathrm{Toffoli}(2)$. $U_\mathrm{s}=I\otimes e^{i\theta_s\sigma_y}\otimes I$ with $\theta_\mathrm{s}=0.282\pi$, $\mathrm{CU}_{\pi/4}^{(2,3)}=I\otimes|0\rangle\langle0|\otimes I+I\otimes|1\rangle\langle1|\otimes e^{i\pi/4\sigma_y}$, $\mathrm{Toffoli}(2)$ is the Toffoli gate with target qubit $Q_2$. It is found that the output fluctuation of $|\psi\rangle_\mathrm{F}$ (green bars) is narrower than the one of $|\varphi\rangle_{\mathrm{c1}}$ (pink bars). Such result can be treated as a time-domain wavefunction localizes in a tightly coupled chain. The introduce of the auxiliary states will increase the probability of the evolving state to return to $|\varphi\rangle_{\mathrm{c1}}$ under the random noise. In Fig. 2(b), the output $P_3(0)$ versus the gate infidelity $1-\bar{F}$ is investigated. The green-solid line with circles: fuzzy input $|\psi\rangle_\mathrm{F}$; The red-dashed line with squares: single input $|\psi\rangle_\mathrm{c1}$. The green zone symbols the fluctuation of $|\psi\rangle_\mathrm{F}$ while the pink zone symbols the ones of $|\psi\rangle_\mathrm{c1}$. The data has been averaged for 1000 times. As can be seen that, the fluctuation of $|\psi\rangle_\mathrm{F}$ will be significantly smaller than the one of $|\psi\rangle_\mathrm{c1}$ in all regions, especially when the infidelity increase. Furthermore, $P_3{(0)}$ of $|\psi\rangle_\mathrm{F}$ declines more slowly than the one of $|\psi\rangle_\mathrm{c1}$ as the infidelity increase.

\begin{figure}[ptb]
\begin{center}
\includegraphics[width=8.5cm]{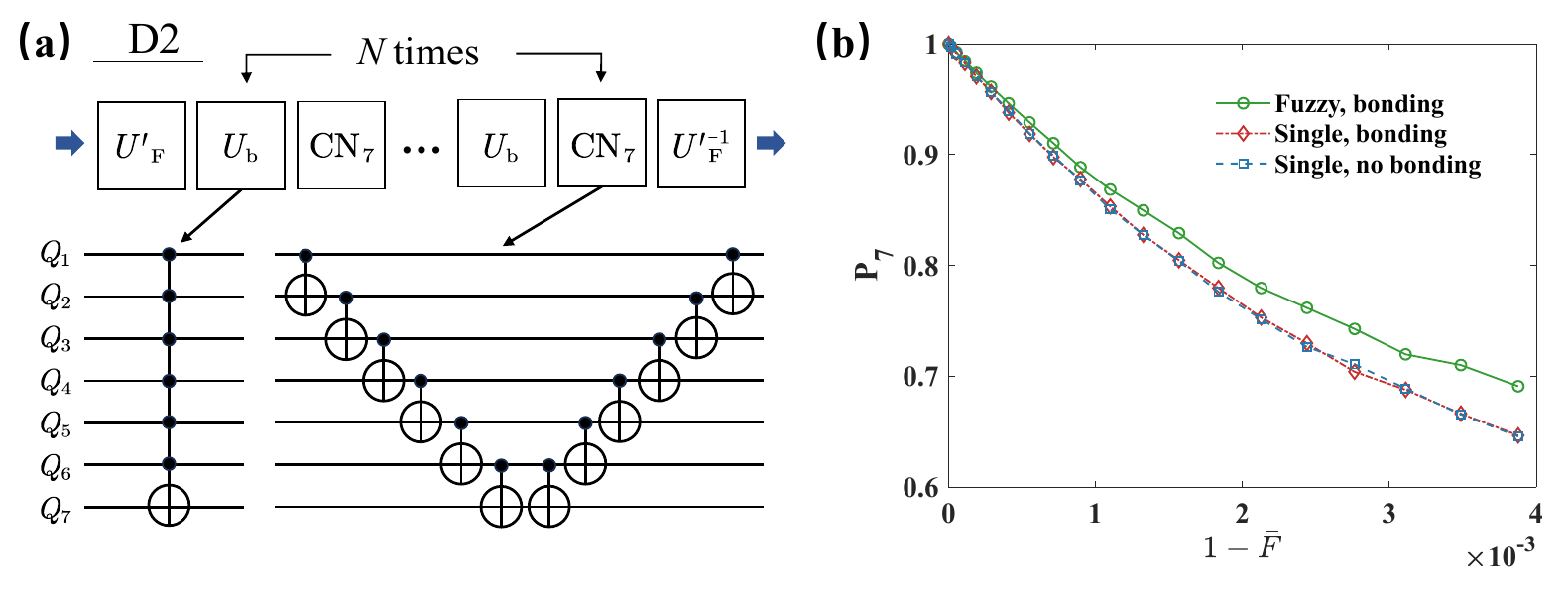}
\caption{
Fuzzy encoding with bonding layers. (a) The tested circuit D2. (b) Performance with different input and operations. The results are averaged by 500 times while the errorbars have nor been displayed to avoid affecting the data presentation.
}
\end{center}
\end{figure}

It is obvious that the suppression of the output distribution requires the qubits to be tightly coupled. In the following we demonstrate how to handle the circuits with loosely coupled qubits. We introduce quantity $C$ to characterize the correlation between the qubits which is defined as $C=\sum_{i=1}^{N_q-1}\sum_{j=i+1}^{N_q}\mathrm{Corr}(T_i,T_j)/\mathrm{C}^2_{N_q}$. $T_i=N_{ti}/N_2$, $N_{ti}$ is the numbers of two-qubit gates that related to qubit $i$, $N_2$ is the number of two-qubit gates in the circuit. $\mathrm{Corr}(T_i,T_j)=e^{-10\times(T_i-T_j)^2}$ characterizes the difference between $i$ and $j$. $\mathrm{C}^2_{N_q}$ is the combination number describing the number of all possible two-qubit gates among $N_q$ qubits. For instance, $C=1$ for the case of circuit D1. In Fig. 3(a), a quantum circuit D2 is introduced through combining the bonding layer ($U_\mathrm{b}$) and the $7$-qubit Toffoli gate ($\mathrm{CN_7}$) $N$ times. In the simulation of circuit D2, the $7$-qubit Toffoli gate is decomposition into 127 elementary quantum gates (2 single-qubit gates and 125 two-qubit gates). $U_\mathrm{b}$ is defined as
\begin{equation}
U_\mathrm{b}=\prod\limits_{i=1}^6\mathrm{CN_2}^{(i,i+1)}\prod\limits_{i=7}^2\mathrm{CN_2}^{(i,i-1)},
\end{equation}
$\mathrm{CN_2}^{(i,j)}$ is the controlled-NOT gate between qubit $i$ and $j$. In the ideal case, $U_\mathrm{b}$ is the identity matrix. However, due to the actual imperfect quantum gates, $U_\mathrm{b}$ will not be the identity matrix and will cause transition between different computational basis. For a single modular $\mathrm{CN_7}$, the correlation $C=0.671$ $(N_q=7, N_2=125)$. When combining $\mathrm{CN_7}$ and $U_b$, the correlation increases to $C=0.724$ $(N_q=7, N_2=137)$. Therefore, the introduce of the bonding layers will enhance the correlations between the qubits. In Fig. 3(b), the output of $Q_7$ ($P_7$) in circuit D2 versus infidelity $1-\bar{F}$ is shown, with $N=10$. The control errors are introduced by
\begin{equation}
D'_2(i)=D_2(i)\bigotimes\limits_{k=1}^7R_i^{(k)}.
\end{equation}
$D_2(i)$ is the $i$-th computational layer of D2 and $R_i^{(k)}$ is the error acting on $k$-th qubit of the $i$-th layer, $R_i^{(k)}=e^{i\eta_k\sigma^{(k)}_r}$, $\sigma^{(k)}_r$ is one of the Pauli matrix with random-choice, $\eta_k/\pi\in[0,0.008]$. For the single state input $|1\rangle^{\otimes 7}$, the evolution operation is $U_{D_2}=(\mathrm{CN_7})^{10}$ ($(U_\mathrm{b}\mathrm{CN_7})^{10}$ for the case with bonding layers). The fuzzy state can be prepared by $|\psi\rangle_\mathrm{F}=U'_\mathrm{F}|1\rangle^{\otimes 7}=\widetilde{|1\rangle^{\otimes 7}}=\sqrt{0.6}|1\rangle^{\otimes 7}+\sqrt{0.2}|1\rangle^{\otimes 4}|110\rangle+\sqrt{0.2}|1\rangle^{\otimes 4}|101\rangle$, $U'_\mathrm{F}=(I^{\otimes 4})U_\mathrm{F}$. The evolution operation of $|\psi\rangle_\mathrm{F}$ is given by $U_{D_2}=U'_\mathrm{F}(U_\mathrm{b}\mathrm{CN_7})^{10}{U'}_\mathrm{F}^{-1}$, ${U'}_\mathrm{F}^{-1}$ is the inverse operation of $U'_\mathrm{F}$ which is used to defuzzify the states \cite{Note2}. The simulation results in Fig. 3(b) has been averaged for 500 times. The green-solid line with circles: fuzzy input with bonding layers; red dashed-dotted line with diamonds: single input with bonding layers; blue-dashed line with squares: single input without bonding layers. As seen, the case of fuzzy input with bonding layer performs best in the three cases. Therefore, the fuzzy encoding of the qubits is an effective way to improve the performance of thousand-qubit gate-level quantum circuits.

\begin{figure}[ptb]
\begin{center}
\includegraphics[width=8.5cm]{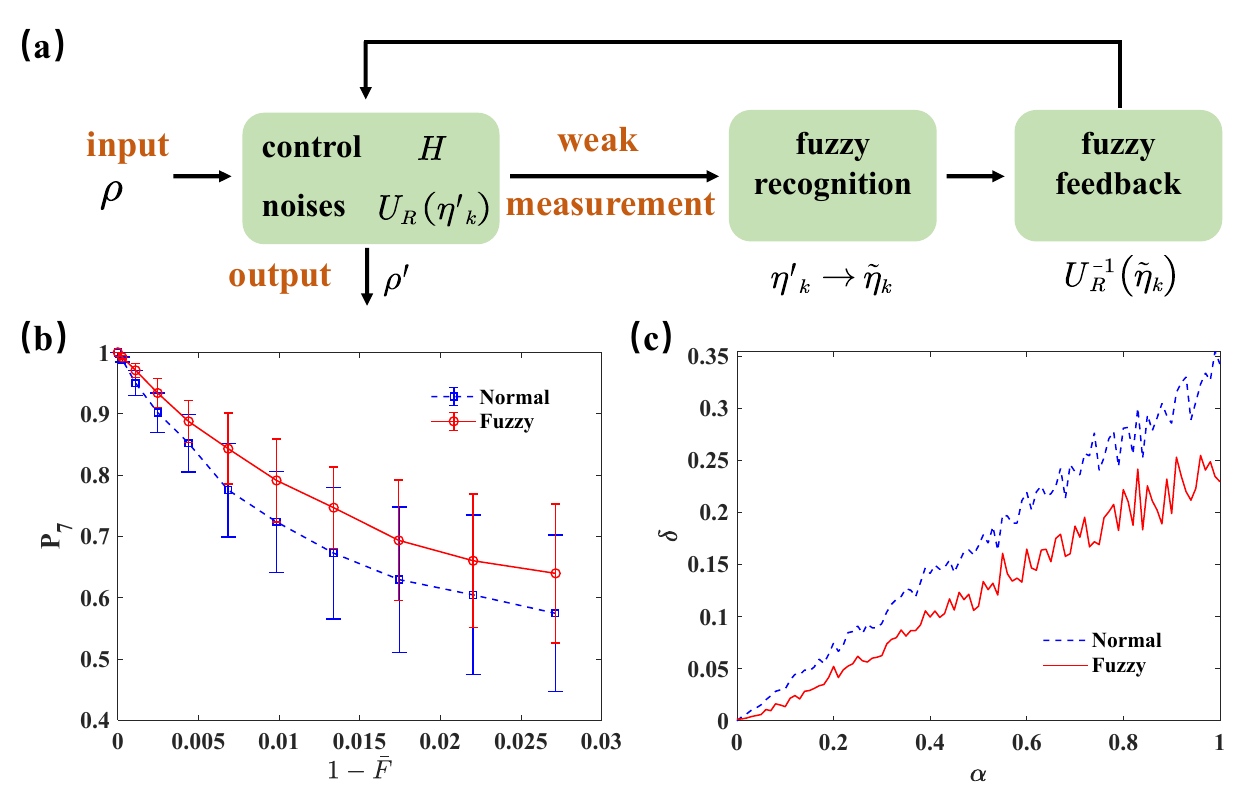}
\caption{
Investigation of quantum control with fuzzy feedback. (a) Schematic diagram of quantum control with fuzzy feedback. (b) The output $P_7$ versus infidelity with different means of feedback. (c) The standard deviation of the estimation of the errors versus the measurement uncertainty $\alpha$.
}
\end{center}
\end{figure}

\textit{Fuzzy feedback of qubits.}--- As depicted in Fig. 4(a), the fuzzy feedback that applied to qubits is investigated. The core idea is to extract the error information from the system evolution through continuous weak measurements. Basing on fuzzy processing, effective long-term feedback on the system can be achieved even in the absence of precise information of the qubits. The continuous weak measurement is an effective means to assist quantum control, which has been used in long coherence time qubits control, quantum steering and QEC \cite{Vijay2012,Murch2013,Livingston2022,Berritta2024}. When the probe field is extremely weak and the dispersion is very large, the measurement induced decoherence and the backaction can be neglected, at a price of increasing the uncertainty of the states information. In this case, the evolution of density matrix $\rho$ with quantum feedback can be described as $\rho'=U^\dagger_{\mathrm{FB}}U^\dagger\rho UU_{\mathrm{FB}}$. $U=U_\mathrm{H}U_\mathrm{R}(\eta_k, \sigma^{(k)}_r)$ is the actual evolution operator and $U_\mathrm{H}$ is the ideal one, where $U_\mathrm{R}(\eta_k, \sigma^{(k)}_r)=\bigotimes\limits_kR_i^{(k)}$ are the errors introduced in Eq. (7). $U_\mathrm{\mathrm{FB}}$ is the imperfect feedback operation that due to the imperfect measurements.

Assuming that the estimation of error-type $\sigma^{(k)}_r$ is correct while the one of real parameter $\eta_k$ is biased, the feedback operation will be given by $U_\mathrm{FB}=U_\mathrm{R}^{-1}(\eta'_k, \sigma^{(k)}_r)=e^{-i\eta'_k\sigma^{(k)}_r}$, $\eta'_k$ is the direct measurement result of $\eta_k$. To realize fuzzy feedback, $\eta'_k$ needs to be transformed to a fuzzy value $\widetilde{\eta}'_k$. It is first to introduce a membership function $\mathcal{V}_j$ to characterize the degree to which $\eta'_k$ belongs to $\eta_j$ with $\mathcal{V}_j=\exp[-(\eta_j-\eta'_k)^2/\xi^2]$, $\eta/(2\eta_0)=(\eta_j)/(2\eta_0)=-0.5:0.05:0.5$, $\xi=0.05\eta_0$. Then the fuzzy vector associated with $\eta'_k$ will be given by $\mathcal{V}=(\mathcal{V}_j)$. As aforementioned in Eq.(1), a fuzzy matrix is also introduced to establish input-output transformation as given by
\begin{equation}
\mathcal{M}_{ij}=\exp[-(\eta_i-\eta_j)^2/\xi^2].
\end{equation}
The output fuzzy vector will be derived as $\mathcal{V}'=\mathcal{M}\circ \mathcal{V}$. Finally, the estimation of $\eta'_k$ according to fuzzy mathematic will be given by $\widetilde{\eta}'_k=\mathcal{V}'\eta$. The fuzzy feedback operation will be given by $\widetilde{U}_\mathrm{\mathrm{FD}}=e^{-i\widetilde{\eta}'_k\sigma^{(k)}_r}$. Replacing $U_\mathrm{FB}$ by $\widetilde{U}_\mathrm{FB}$, the fuzzy feedback is realized.

In Fig. 4(b), the results of quantum circuit using fuzzy feedback are shown. Here we adopt circuit D2 with $N=10$, $\eta_0=0.02\pi$. The direct obtained results are modelled by $\eta'_k=(1+\alpha rand)\eta_k$, $\alpha$ characterize the measurement uncertainty. We have chosen $\alpha=1$ which is a rather bad case. The input is fuzzy state $\widetilde{|1\rangle^{\otimes 7}}$ and the data has been averaged for 100 times. The red-solid line with circles represents the results fuzzy control $\widetilde{U}_\mathrm{FB}$. $P_7$ is still closed to 0.7 when the infidelity is larger than 0.025, which is a significant improvement. The blue-dashed line with squares shows the results with feedback operation $U_{\mathrm{FB}}$. As can be seen that, the performance of feedback with fuzzy mathematics is much better than the one only depends on the direct measurement results. To understand Fig. 4(b), we introduce the standard deviation $\delta(K)=\sum\sqrt{(K-\eta_k)^2}/N_q, K=\eta'_k, \widetilde{\eta}'_k$, the summation runs over the $N_q$ qubits. The comparison between fuzzy feedback and direct feedback is shown in Fig. 4(c). The fuzzy mathematics gives a better estimation of the real parameters based on the measurement results, which makes the feedback more effectively. With a modest measurement precision $\alpha=0.5$ and fuzzy feedback in circuit D2, one can run $180,000$ level quantum gates with an average fidelity 0.999, which is sufficient for many useful algorithms \cite{Preskill2018}.

\begin{figure}[ptb]
\begin{center}
\includegraphics[width=8.5cm]{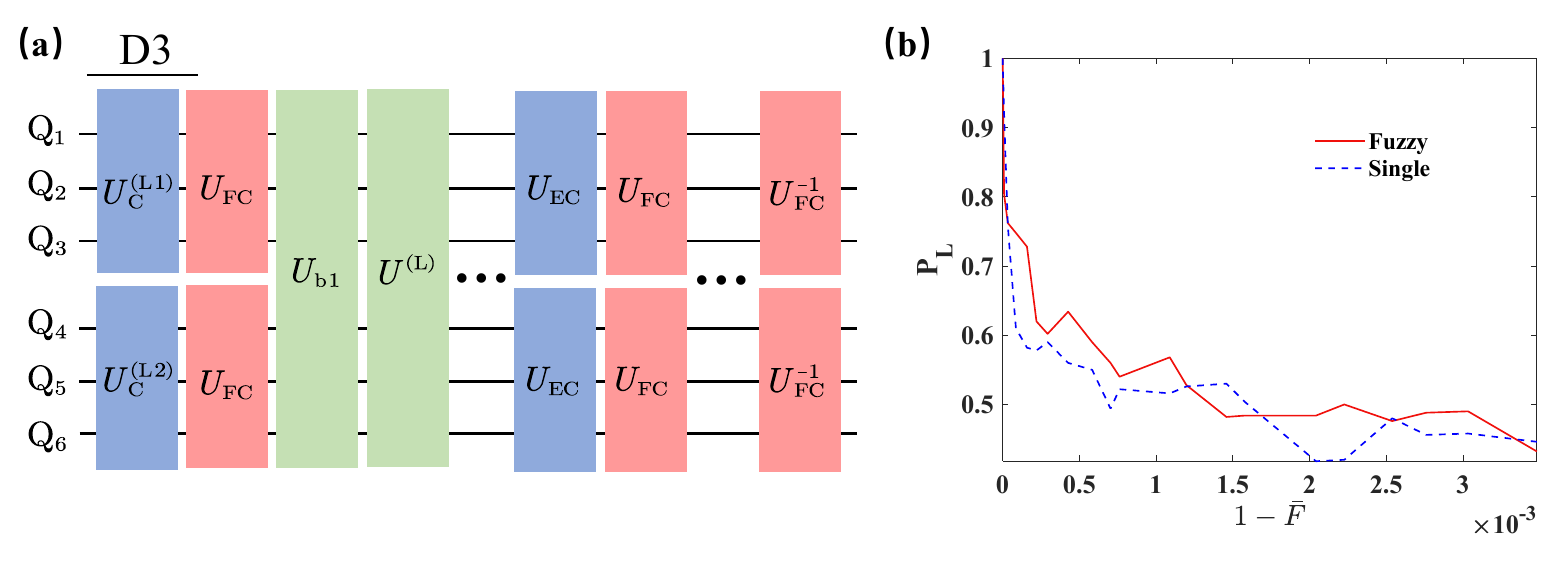}
\caption{
Incorporation of fuzzy encoding and QEC. (a) The tested circuit D3. (b) The output $P_\mathrm{L}$ versus infidelity $1-\bar{F}$ with fuzzy state input and single state input, respectively.
}
\end{center}
\end{figure}

\textit{Combining fuzzy encoding and QEC.}--- Finally, we want to emphasize that the fuzzy processing can be integrated into QEC. In the circuit D3 in Fig. 5(a), the incorporation of fuzzy encoding and QEC is shown, where $\{Q_1, Q_2, Q_3\}$ are the control qubits and $\{Q_4, Q_5, Q_6\}$ are the target qubits. As a demonstration, we consider the correction of flip errors. The QEC encoding operations will be given by $U^{\mathrm{(L1)}}_{\mathrm{C}}={\mathrm{CN_2}^{(1,2)}}\otimes {\mathrm{CN_2}^{(1,3)}}$ and $U^{\mathrm{(L2)}}_{\mathrm{C}}={\mathrm{CN_2}^{(4,5)}}\otimes {\mathrm{CN_2}^{(4,6)}}$ and the logical states will be defined as $|1\rangle_{\mathrm{L}}=|111\rangle$ and $|0\rangle_{\mathrm{L}}=|000\rangle$. For instance, by choosing $|\psi\rangle_1=|\psi\rangle_4=|1\rangle$, the two-qubit logical state will be derived as $|\psi\rangle^\mathrm{(L)}_\mathrm{in}=|1\rangle_{\mathrm{L}}\otimes|1\rangle_{\mathrm{L}}=|111\rangle\otimes|111\rangle$ after the encoding. In Fig. 5(a), the logical operation in the quantum circuit is introduced as $U^{(L)}=(U_u^{(L)}U_d^{(L)})^N$, where the transverse gates are adopted to realize error resilient gates with $U_u^{(L)}={\mathrm{CN_2}^{(1,4)}}\otimes {\mathrm{CN_2}^{(2,5)}}\otimes {\mathrm{CN_2}^{(3,6)}}$ and $U_d^{(L)}={\mathrm{CN_2}^{(4,1)}}\otimes {\mathrm{CN_2}^{(5,2)}}\otimes {\mathrm{CN_2}^{(6,3)}}$. After $N_\mathrm{t}$ logical operations, the logical states should be detected and the error states should be flipped to one of the logical states through operations $U_{\mathrm{EC}}$ ($N_\mathrm{t}=1$ for the explicit quantum error correction). The final state will be given by $|\psi\rangle^\mathrm{(L)}_\mathrm{out}=\sum_k\beta_k|k\rangle_L, k=00, 01, 10, 11$ and the decoding state is $|\psi\rangle_\mathrm{out}=\sum_k\beta_k|k\rangle$. We focus on the population of target qubit which is $P_L=(1+\mathbf{Tr}(\rho_\mathrm{t}\sigma_z))/2$, $\rho_t=\mathbf{Tr}_\mathrm{c}(\rho_{\mathrm{out}})$ is the reduced density matrix of target qubit after tracing by the control qubit state, $\rho_{\mathrm{out}}=|\psi\rangle_\mathrm{out}\langle\psi|$.

To generalize the above processing to the case of fuzzy processing, the fuzzy input is introduced $|\widetilde{\psi}\rangle^\mathrm{(L)}_{\mathrm{in}}=|\widetilde{111}\rangle\otimes|\widetilde{111}\rangle$, here the fuzzy state preparation is $U_\mathrm{FC}=U_{\mathrm{F}}$. Furthermore, to increase the correlation between the qubits, the bonding operation is added with $U_{b1}=\prod\limits_{i=1}^5\mathrm{CN_2}^{(i,i+1)}\prod\limits_{i=6}^2\mathrm{CN_2}^{(i,i-1)}$. The logical operation for the fuzzy processing will be given by $\widetilde{U}^{(L)}=(U_{b1}U_u^{(L)}U_d^{(L)})^N$. The measurement of target qubit here is similar to the case of single state input $|\psi\rangle^\mathrm{(L)}_\mathrm{in}$, with an additional defuzzy operation $U^{-1}_{\mathrm{FC}}$.

In Fig. 5(b), we compare the performance of QEC with $|\widetilde{\psi}\rangle^\mathrm{(L)}_{\mathrm{in}}$ and $|\psi\rangle^\mathrm{(L)}_\mathrm{in}$ input, respectively. The random noise is introduced to each logical operation with ${U'}_k^{(L)}=U_k^{(L)}\bigotimes\limits_{l=1}^2R_k^{(l)}$, here we assume that only one error happens in the physical qubit $k$ of logical qubit $l$ with $R_k^{(l)}=e^{if(\eta_k)\sigma_x}$. $f(\eta_k)=\beta\pi/2[p(\pi/2)+p(\pi/50)/25+p(\pi/500)/250]$ characterize the probability of returning large errors, moderate errors, and small errors. $\eta_kp(\eta_k)$ return $\eta_k$ with probability $p(\eta_k)$. $\beta\in[0,0.2]$ describes the noise intensity. Without loss of generality, we set $p(\pi/2)=0.006, p(\pi/50)=0.006, p(\pi/500)=0.15$. $N_t=2000$ is adopted which means that the system is measured and being corrected every 2000 logical operations and we test the output with 18 rounds QEC ($N=18N_t=36000$ logical operations), namely, $3N\sim10^5$ level physical gates is included in circuit D3. As shown in Fig. 5(b), the output $P_\mathrm{L}$ versus infidelity $1-\bar{F}$ with different input is shown which is averaged by 500 times. In the ideal case, $P_\mathrm{L}=1\ (\beta=0)$. As the noise intensity $\beta$ and also the infidelity increase, $P_7$ declines. Nevertheless, $P_7$ of $|\widetilde{\psi}\rangle^\mathrm{(L)}_{\mathrm{in}}$ (red-solid line) declines more slowly than the case of $|\psi\rangle^\mathrm{(L)}_\mathrm{in}$ (blue-dashed line). When the infidelity is larger than $10^{-3}$, the output for the both cases approach $0.5$, producing inconclusive results. To further improve the performance, one may increase the number of physical qubits $N_{\mathrm{EC}}$ that encoding logical qubits. The permitted logical operation will be increased as $(p_\mathrm{c})^{-(d+1)/2}$, $p_\mathrm{c}$ is the error rate of physical qubits and $d\approx N^2_{\mathrm{EC}}$ is the encoding distance.

\textit{Conclusion.}--- In summary, we have proposed a scheme to realize quantum computation with fuzzy processing, which is constructed by fuzzy recognition, fuzzy encoding and fuzzy feedback. The fuzzy processing of qubit states would reduce the required gate fidelity, epsically for the two-qubit control, reducing the difficulty of implementation. The fuzzy processing combined with QEC abandon the requirement of timely measurement and feedback of quantum states. The proposed scheme will be fairly suitable for the solution of decision problems, which has important use in the optimization problems and control problems. The update of the framework of quantum computations will facilitate the discovery of more interesting quantum algorithms.

\textit{Acknowledgment.}--- This work was supported by Natural Science Foundation of Guangdong province (Grant No. 2024A1515012516), Guangdong Provincial Quantum Science Strategic Initiative (Grant No. GDZX2303006).


\begin{thebibliography}{99}
\bibitem{limit}The limit of the data has been set to 0.5 when the infidelity is large.
\end{thebibliography}
\end{document}